\documentclass{article}

\usepackage[usenames,dvipsnames]{xcolor}

\usepackage{cite}
\usepackage{caption}
\usepackage{graphicx}
\usepackage{epsfig}
\usepackage{amsmath,amssymb,bm,stmaryrd}
\usepackage{delarray}
\usepackage{stfloats}
\usepackage{float}
\usepackage{multirow,hhline}

\usepackage{algorithm}
\usepackage{algorithmic}
\usepackage{arydshln}
\usepackage{cuted}
\usepackage{balance}
\usepackage{verbatim}
\usepackage{array}

\begin{document}

%+Title
\title{Robust Algorithms for Localizing Moving Nodes in Wireless Sensor Networks}

%\author{\begin{normalsize} Hadeel Elayan and Raed M. Shubair\end{normalsize}\\ \begin{center}\begin{tiny}\begin{normalsize} Electrical and Computer Engineering Department, Khalifa University, UAE\end{normalsize}\end{tiny}\end{center} \\\begin{small} Email: hadeel.mohammad@kustar.ac.ae, raed.shubair@kustar.ac.ae\end{scriptsize}}
\author{\begin{normalsize} Hadeel Elayan$^1$ and Raed Shubair$^2$\end{normalsize}\\ \normalsize $^1$Electrical and Computer Engineering Department, Khalifa University, UAE  \\ \normalsize
$^2$Research Laboratory of Electronics (RLE), MIT\\ \small Email: hadeel.mohammad@kustar.ac.ae, rshubair@mit.edu}
\date{2016}
\maketitle
%-Title

%+Abstract

The vivid success of the emerging wireless sensor technology (WSN) gave rise to the notion of localization in the communications field. Indeed, the interest in localization grew further with the proliferation of the wireless sensor network applications including medicine, military as well as transport. By utilizing a subset of sensor terminals, gathered data in a WSN can be both identified and correlated which helps in managing the nodes distributed throughout the network. In most scenarios presented in the literature, the nodes to be localized are often considered static. However, as we are heading towards the $5^{th}$ generation mobile communication, the aspect of mobility should be regarded. Thus, the novelty of this research relies in its ability to merge the robotics as well as WSN fields creating a state of art for the localization of moving nodes. The challenging aspect relies in the capability of merging these two platforms in a way where the limitations of each is minimized as much as possible. A hybrid technique which combines both the Particle Filter (PF) method and the Time Difference of Arrival Technique (TDOA) is presented. Simulation results indicate that the proposed approach outperforms other techniques in terms of accuracy and robustness. 


\section{Introduction}
% no \IEEEPARstart
Localization has
received a considerable amount of attention over the past decade since it is an essential capability enabling other applications \cite{shubair2015vivo}. The localization problem is the process of determining the position of a node relative to a given reference frame of coordinates in a Wireless Sensor Network (WSN). Undeniably, the sensor' location must be known for its data to be meaningful \cite{4407221}. The prominence of this process comprises the ability to identify and correlate gathered data in a wireless sensor network as well as manage nodes located in a determined location. In fact, localization is achieved through the utilization of a subset of sensor terminals \cite{Ilyas:2004:HSN:983182}. Nodes which have known states at all times and are aware of their own global coordinates a priori are called anchor nodes or beacon nodes. The anchor nodes are equipped with a Global Positioning System (GPS) thereby assisting in computing other nodes'' locations by using a number of techniques such as lateration, angulation or a combination of both \cite{mirebrahim2009monte,kulaib2011overview}. Nevertheless, sensor nodes which have a priori unknown states 
and need to determine their positions using a localization
algorithm are referred to as agents  \cite{5450293}.

Several localization algorithms have been proposed in the literature \cite{hightower2001location}. One of the most important aspects of  localization  involves measuring the transmission range of the wireless signal. Range-based localization relies on the availability of point-to-point distance (or angle information). The obtained measurements of different ranging techniques  such as   Time of Arrival (TOA), Time Difference of Arrival (TDOA), Angle of Arrival (AOA), and Received Signal Strength (RSS) are the keys for range-based schemes \cite{wang2010survey,najibi2013localization,shubair2007setup,al2004computationally,al2005computationally,al2003investigation,samhan2006design,al2003performance,shubair2004robust,nwalozie2013simple,al2006direction,al2005direction,shubair2005robust,bakhar2009eigen,shubair2005performance,alhajri2015hybrid}.

Nonetheless, most of the prevailing localization algorithms used in WSNs assume   static nodes which do not move after deployment. Hence, the mobility effect is not explicitly considered in the analysis approach.  The mobility effect cannot be ignored in certain WSN platforms such as intelligent transportation, patient monitoring, and habitual tracking. The major technical challenge in localization of moving nodes is the rapidly changing localization scenarios resulting in inaccurate prediction of node locations \cite{mirebrahim2009monte}. It must be noted as well that mobile networks convey different characteristics from static networks. Other important aspects that deteriorate the performance of mobile sensor networks are issues related to  the changing topology of the network, the varying connectivity, and latency problems \cite{5432182}. These effects stimulate the need for developing robust yet accurate   the localization algorithms for moving nodes in WSNs

The Monte Carlo  localization (MCL) \cite{dellaert1999monte} was the first practical method for localization of mobile WSNs. MCL applies the Sequential Monte Carlo (SMC) method \cite{handschin1970monte} to achieve localization. The reason behind this is that the posterior distribution of a sensor node after movement can be naturally formalized using a nonlinear discrete time model and the SMC method provides simple simulation-based approaches to estimate the distribution \cite{5432182}. Basically, the method assumes no functional form, but instead, it uses a set of random samples (also called particles) to estimate the posteriors. When the particles are properly placed, weighted, and propagated, posteriors can be estimated sequentially over time. This technique is more popularly known as the Particle Filter (PF)\cite{rui2001better}. Nevertheless, previous SMC-based localization algorithms either suffer from low sampling efficiency or require high beacon density to achieve high localization accuracy.

 In this chapter, we discuss the deployment of the Particle Filter (PF) framework into the localization of moving nodes in an open environment  using TDOA measurements. This incorporation accurately captures the mean and covariance which improves the localization accuracy and helps achieve  a robust performance. The rest of the paper is organized as follows. Section II revists the fundamental localization techniques. Section III presents the system model and problem statement. Section IV  describes the  proposed TDOA-PF technique. In Section V, numerical validation on simulated scenarios can be found. Finally, the conclusions are summarized in  Section VI.

\section{Localization Techniques}
Time-of-arrival (TOA), time-difference-of-arrival (TDOA), received signal strength (RSS) and direction-of-arrival (DOA) of the emitted signal are commonly used measurement techniques in WSN localization. Basically, TOAs, TDOAs and RSSs provide the distance information between the source and sensors while DOAs are the source bearings relative to the receivers. Nevertheless, finding the source position is not a trivial task because these measurements have nonlinear relationships with the source position \cite{so2011source}. The signal models and their basic positioning principles are generalized as follows:
\begin{equation}\label{two}
r=f(x)+n\ 
\end{equation}
where $r$ is the measurement vector, $x$ is the source position to be determined, $f(x)$  is a known nonlinear function in $x$, and $n$ is an additive zero-mean noise vector.
\subsection{Time of Arrival}
TOA is defined as the difference between the sending time of the signal at the transmitter and the receiving time of the signal at the receiver (time delay). The time delay can be computed by dividing the separation distance between the nodes by the propagation velocity. TOA technique uses multilateration, since it includes ranges from more than three reference points. Mathematically, the TOA measurement model is formulated as follows. Let $x=[x  y]^T $ be the unknown source position and $x_l=[x_l  y_l]^T$ be the known coordinates of the $l_{th}$ sensor, $ l=1,2,…...,L,$ where $L\geq3$ is the number of receivers. The distance between the source and the, $l_{th}$ sensor, denoted by $d_l$ is:
\begin{equation}\label{x}
d_{l}=[|x-x_{l}|]_{2}=\sqrt{(x-x_{l})^2+(y-y_{l})^2}\ 
\end{equation}
Without loss of generality, we assume that the source emits a signal at time $0$, and the $l_{th}$ sensor receives it at time, $t_{l}$ . That is, ${\{t_l\}}$  are the TOAs; a simple relationship between $t_l$  and $d_l$ is given by:
 
 \begin{equation}\label{y}
t_{l}=\frac{d_{l}}{c}\ 
\end{equation}
where $c$ is the propagation speed of the radio signal (speed of light).
In practice, TOAs are subject to measurement errors. As a result, the range measurement based on multiplying $t_l$  by $c$ denoted by $r_{(TOA,l)}$ is modeled as: 

\begin{equation}\label{x}
r_{TOA,l}=d_{1}+\eta_{TOA,l}=\sqrt{(x-x_{l})^2+(y-y_{l})^2}\ +\eta_{TOA,l}\
\end{equation}
where $\eta_{(TOA,l)}$ is the range error in $r_{(TOA,l)}$ which  results from the TOA disturbance.

In TOA, the nodes have to be synchronized and the signal must include the time stamp information. This requirement adds to the cost of the signal by demanding a highly accurate clock and increasing the complexity of the network. To overcome such restrictions, the RTOA (Round-trip Time of Arrival) and TDOA are introduced \cite{so2011source}.

RTOA is the most practical scheme in decentralized settings since it does not require a common time reference between the nodes.  Actually, it measures the difference between the time when a signal is sent by the sensor and the time when the signal returned by a second sensor is received at the original sensor. Because the same clock is used to compute the round trip propagation time, there is no synchronization problem \cite{laaraiedh2010contributions}. \subsection{Time Difference of Arrival}
The basic idea of TDOA is to determine the relative position of the mobile transmitter by examining the difference in time at which the signal arrives at a pair of sensors. This implies that clock synchronization across all receivers is required. Nonetheless, the TDOA scheme is simpler than the TOA method because the latter needs the source to be synchronized as well. Similar to the TOA, multiplying the TDOA by the known propagation speed leads to the range difference between the source and two receivers \cite{so2011source}.

The TDOA measurement model is mathematically formulated as follows. We assume that the source emits a signal at the unknown time $t_{0}$ and the $l_{th}$ sensor receives it at time $t_{l}$,  $l=1,2,...,L$ with $L\geq3$. There are $\frac{L(L-1)}{2}$ distinct TDOAs from all possible sensor pairs, denoted by $t_{k,l}=(t_{k}-t_{0})-(t_{1}-t_{0})$, $k$, $l=1,2,...,L.$ Taking $L=3$ as an example, the distinct TDOAs are $t_{2,1}$ , $t_{3,1}$ and $t_{3,2}$. We easily observe that $t_{3,2}$ = $t_{3,1}$- $t_{2,1}$, which is redundant. In order to reduce complexity without sacrificing estimation performance, we should measure all $\frac{L(L-1)}{2}$ TDOAs and convert them to $(L - 1)$ non-redundant TDOAs for source localization \cite{4524057}. We consider the first sensor as the reference and the non-redundant TDOAs are $t_{(l,1)}$, $l=1,2,…,L.$ The range difference measurements deduced from the TDOAs are modeled as: 
\begin{equation}\label{x}
r_{TDOA,l}=d_{1}+\eta_{TDOA,l}=\sqrt{(x-x_{l})^2+(y-y_{l})^2}\ +\eta_{TDOA,l}\
\end{equation}

where

\begin{equation}\label{x}
d_{l,1}=d_{l}+d_{1}\
\end{equation}
and $\eta_{(TDOA,l)}$ is the range error in $r_{(TDOA,l)}$ which is proportional to the disturbance in $t_{(l,1)}$.

\subsection{Received Signal Strength}
The RSS approach is used to estimate the distance between two nodes based on the strength of the signal received by another node. A sender node sends a signal with a determined strength that fades as the signal propagates. It is known that the bigger the distance to the receiver node, the lower the signal strength when it arrives to the node.  In fact, the node can calculate its distance from the transmitter using the power of the received signal, knowledge of the transmitted power, and the path-loss model. The operation starts when an anchor node broadcasts a signal that is received by the transceiver circuitry and passed to the Received Signal Strength Indicator (RSSI) to determine the power of the received signal [9]. 

By assuming that the source transmitted power is $P_{t}$  and there is no disturbance, the average power received at the $l_{th}$ sensor, denoted by $P_{(r,l)}$ is modeled as:

\begin{equation}\label{x}
P_{r,l}=K_{l}P_{t}d^{-\alpha}\
\end{equation}
where $K_l$  accounts for all other factors which affect the received power, including the antenna height and antenna gain, while ? is the path loss constant. Actually, the value of $\alpha$ can vary between 2 and 5. Particularly,  $\alpha =2$ depicts free space. 

\subsection{Direction of Arrival}
The DOA estimation method requires the base stations to have multiple antenna arrays for measuring the arrival angles of the transmitted signal from the mobile station at the base stations. This technique can be further divided into two subclasses, those making use of the receiver antenna's amplitude response and those making use of the receiver antenna phase response. The accuracy of the DOA measurements is limited by the directivity of the antenna, shadowing and multipath reflections \cite{boudhir2010new}. Although this scheme does not require clock synchronization as in RSS-based positioning, an antenna array is needed to be installed at each receiver for DOA estimation. Assuming $\phi_{1}$ be the DOA between the source and $l_{th}$ receiver, we have
\begin{equation}\label{z}
tan({\phi_{1}})=\frac{y-y_{l}}{x-x_{l}}, l=1,2,...,L\
\end{equation}

with $L\geq2$. Actually, $\phi_{1}$ is the angle between the line-of-bearing from the $l_{th}$ receiver to the target and the x-axis. The DOA in the presence of angle errors, denoted by $\{r_{(DOA,l)}\}$, are modeled as: 

\begin{equation}\label{z}
r_{(DOA,l)}=\phi_{1}+\eta_{(DOA,l)}=tan^{-1}({\frac{y-y_{l}}{x-x_{l}}})+\eta_{(DOA,l)}\
\end{equation}
where $\{{\eta_{(DOA,l)}}\}$ are the noises in $\{r_{(DOA,l)}\}$ which are assumed zero-mean uncorrelated Gaussian processes. 
 
\section{System Model and Problem Statement}
\subsection{Problem Statement}
The problem to be investigated can be viewed in Figure 1 and is described as follows. Basically, it consists of a mobile robot carrying a sensor node and of multiple anchor nodes. We refer to the mobile robot as a mobile node. Anchors whose locations are known ping out signals which are differentiable from the others. Signals emitted from multiple anchors are used by the mobile node to determine its location. Although the mobile node recognizes the time of arrival of the pings, it has no information on the time of ping emission. Hence, it can use only the difference of arrival times of the pings to achieve localization. The following are the notations used for the solution:
\begin{itemize}
\item 
$d_{i}$ = The distance between the moving node and the \textit{i}-th anchor
\item 
$d_{i1}$= $d_{i}$-$d_{1}$: range difference to the \textit{i}-th anchor and reference anchor
\item
$R_{i}$ = distance between the \emph{i}-th anchor and the reference anchor
\item
$R_{r}$ = distance between the moving node and the reference anchor
\end{itemize} 
\begin{figure}[h]
\centering
\includegraphics[width=0.6\textwidth,clip]{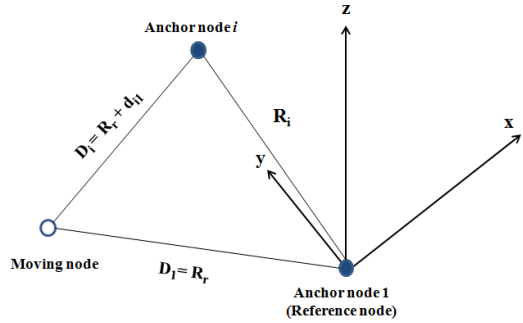}
\caption{The representation of the localization problem where the moving node depends on signals emitted from the anchors to estimate its location.}
\label{system}
\end{figure} 
\subsection{System Model}
\begin{enumerate}
\item \textit{Mobility Model}

 Time is divided into equal segments, $\Delta T$, in which the
mobile node moves along the direction at a constant value as depicted in
Figure 2. The movement is divided into several sub-paths
according to the  time segment where the mobile node progresses at a constant velocity as presented in Figure 3. After using a time segment, the mobile node can change the moving velocity thereby satisfying the uniform distribution in $[Vmin, Vmax]$, as shown in \eqref{one}.

\begin{equation}\label{one}
V\sim\ U [V_{min}, V_{max}]
\end{equation}
\begin{figure}[h]
\centering
\includegraphics[width=0.6\textwidth,clip]{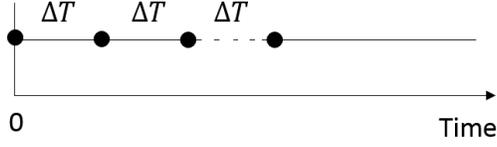}

\caption{Equal Time Segment.}
\label{system}
\end{figure}

\begin{figure}[h]
\centering
\includegraphics[width=0.6\textwidth,clip]{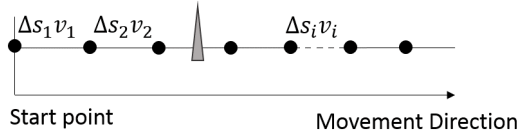}
\caption{Mobility model of a robot.}
\label{system}
\end{figure}
 In Figure 3 above, the
triangle denotes the mobile node, $\Delta s_i$ denotes the
movement distance at time segment $i$ by velocity $v_{i}$. 

\

\item  \textit{Time Difference of Arrival Model}

TDOA is the difference in the time of   the transmitted signal from the unknown sensor at a pair of anchors as explained in the section above. If the first anchor is assigned as a reference point, the range measurements based on the TDOAs are of the form \cite{cheung2006constrained}
\begin{align*}
r_{TDOA,i}=(d_{i}-d_{1})+ \eta _{ TDOA,i}\ 
\end{align*}
\begin{equation}\label{tdoa}
\begin{aligned}
=\sqrt{(x-x_{i})^2+(y-y_{i})^2}-\\\sqrt{(x-x_{i})^2+(y-y_{i})^2}\\ +\eta_{TDOA,i}
\end{aligned}
\end{equation}
where $i=2, 3, ....M$
  and the range error $\eta_{TDOA,i}$ can be obtained from the difference of two TOA noise. So, $\eta_{TDOA,i}$ is $\eta_{TOA,i}- \eta_{TOA,i-1},$ $i=2, 3, ....M.$ Either linear or nonlinear algorithms may be deployed to achieve source localization. These algorithms basically involve minimizing the Least Squares (LS) or Weighted Least Squares (WLS) cost function. Another alternative involves analytically solving the problem through a closed form equation \cite{najibi2013localization}.  \\  

\item \textit{System State Model}

The system state model for the mobile wireless sensor is represented by the following formula.
\begin{equation}\label{two}
x_{k}=x_{k-1}+v_{k}\cdot\Delta T+\epsilon_{k}\ 
\end{equation}

where $x_{k}$ is the position of a mobile node from the
anchor, $\Delta T$ is the time segment,  $v_{k}$ is the current velocity, and $\epsilon_{k}$ denotes the system state noise which obeys Gaussian distribution with zero mean and covariance $Q_{k}$, $R_{k}$. The TDOA measurement is provided in \eqref{tdoa}.
\end{enumerate}
\subsection{System Assumption}
To simplify the problem, certain assumptions need to be 
made.

\begin{itemize}
\item 
 All the sensor nodes own equal physical parameters.
 \item
  The movement velocity of mobile node stays the same at time $\Delta T$, and the time of turning is ignored.
\item
The random velocity obeys normal uniform distribution
as \eqref{one}.
\item
 The anchors are deployed at determined pattern and the
position cannot be changed.

\end{itemize}

\section{ Localization of Moving Nodes Based on PF-TDOA Approach}

 Localizing a moving node is an interesting challenge due to the fact that it demands extensive computational effort at each iteration. To localize a moving sensor node, filtering methods are often applied. Kalman filter is a celebrated technique for recursive state estimation, and has been widely used in many applications. Due to its unbiased minimum variance estimation with white noise assumption, Kalman filter is an optimal solution in linear systems. However, it is limited in non-linear and non-Gaussian systems, which exist ubiquitously in the real world \cite{zhou2011tracking}. To address these limitations, a number of methodologies have been proposed, e.g., Extended Kalman Filter (EKF) \cite{1098986} and Unscented Kalman Filter (UKF) \cite{882463}. Nevertheless, another alternative to the  methods mentioned above is the Particle Filter (PF), which is a Bayesian inference based scheme.

 In recent years, PF has become a research
hotspot in the field of nonlinear filtering and estimation. The
key idea \cite{978374} is to represent the required posterior probability
density function (PDF) by a set of random samples (called
particles) with associated weights and to compute estimates
based on these samples and weights. As the number of
particles increases, the Monte Carlo characterization becomes an equivalent representation to the usual functional description of the posterior PDF, and the PF approaches the optimal Bayesian estimator. 

In order to formulate the PF scheme, assume the prior conditional probability density of a dynamic system is $p(x_{0})$ and let $ \left\{ x^{i}_{0:k}, w^{i}_{k} \right\}^{N}_{i=1}$ represent
both the measurement value and its weight of random sample during
the time $k$. The PDF
is $p(x_{0:k}|z_{1:k}).$ The weights of $x^{i}_{k}$ are normalized such
that $\sum_{i=1}^{N_{x}}w^{i}_{k}=1$. Then, the posterior density at time $k$ can be
approximated as 

\begin{equation}\label{four}
p(x_{0:k}|z_{1:k})\approx\ \sum_{i=1}^{N_{x}}w^{i}_{k}\delta(x_{0:k}-x^{i}_{0:k})
\end{equation}
Following the derivation presented in \cite{978374}, 
we get
\begin{equation}\label{five}
p(x_{k}|z_{1:k})\approx\ \sum_{i=1}^{N}w^{i}_{k}\delta(x_{k}-x^{i}_{k})
\end{equation}
The approximation of $x$ yields 
\begin{equation}\label{six}
\hat x _{k}\approx\ \sum_{i=1}^{N}w^{i }_{k} x^{i}_{k}
\end{equation}
Next, we must compute an estimate of the effective number of particles as
\begin{equation}\label{five}
\hat N_{eff}= \frac{1}{\sum_{i=1}^{N_{x}}(w^{i}_{k})^{2}}
\end{equation}
Actually, if the effective number of particles is less than a given threshold, $\hat N _{eff}< N _{thr}$, resampling must be performed.

The PF algorithm can be used to address  mobile node localization by using time difference of arrival (TDOA). This approach was first presented in  \cite{6263573} for underwater localization. In this paper, we revisit the approach and extend it for the localization of moving nodes in the case of an open environment. In fact, the incorporation between the PF and TDOA makes use of  both the internal motion information of the moving node and the interaction between the distance estimates of the anchor sensors thereby resulting in accurate localization of the moving  node. Actually, localization based on time difference of arrival (TDOA) has turned out to be a promising approach when neither receiver positions nor the positions of signal origins are known a priori. Thus, the PF-TDOA technique  is suitable for TDOA
localization as the state can be computed online and it is
robust to motion and measurement uncertainty \cite{6817863}.

This PF approach actually predicts the state of interest and then corrects the prediction based on the observed sensor data. The prediction and correction steps are repeated as the state progresses with time \cite{1236770}.  Each particle in the PF represents a state which translates into the moving node location in the case of TDOA localization.

The pseudo code of the PF can be viewed in Figure 4.

\begin{figure}[h]
\centering
\includegraphics[width=0.6\textwidth,clip]{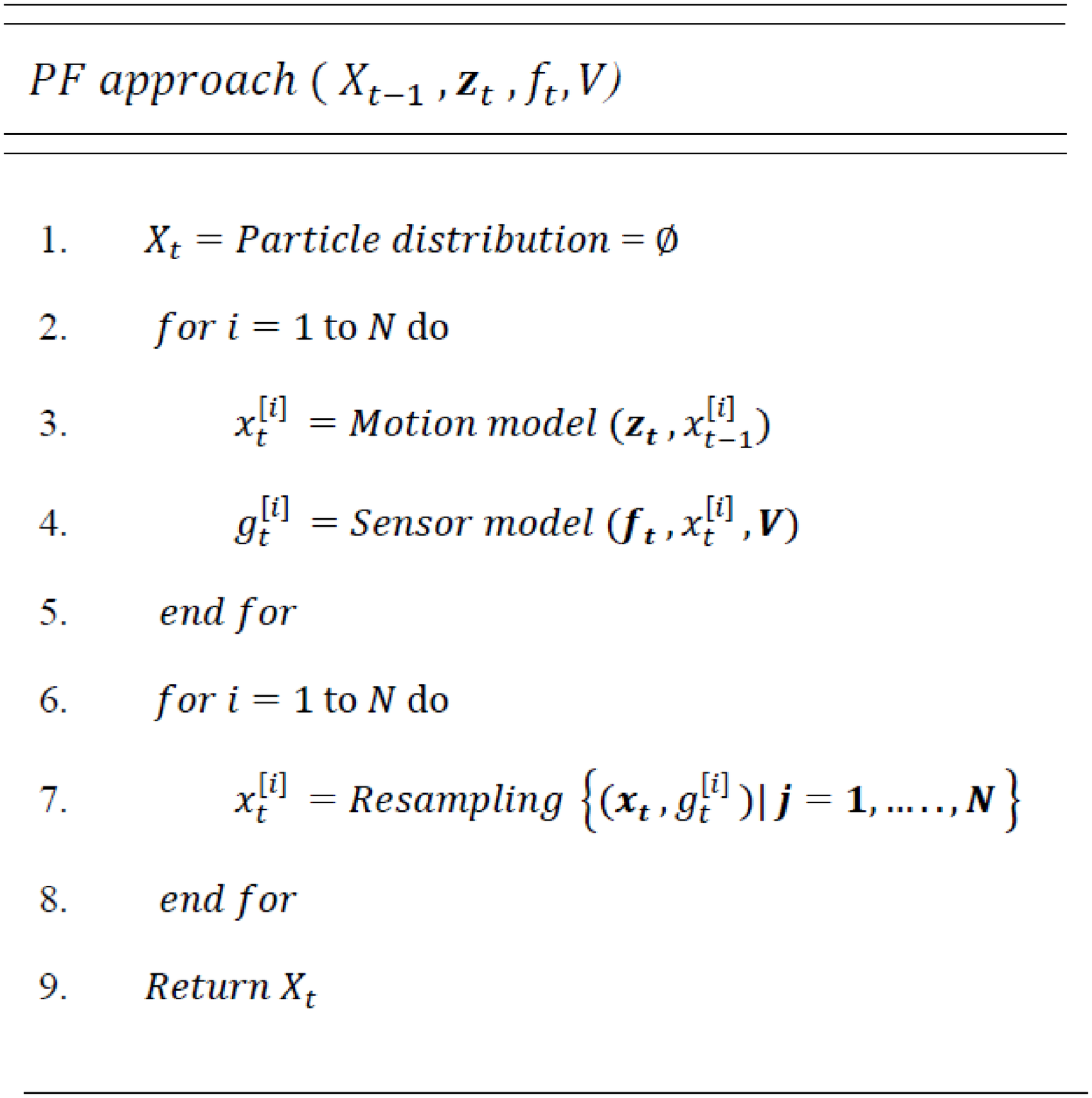}
\caption{PF procedure with number of particles N.}
\label{system}
\end{figure}

The code takes the following as inputs: the particle distribution at the previous sampling time $X_{t-1}$, the observed sensor data $f_{t}$, the internal moving node information $z_t$, the information on the surrounding local environment $V$ that affects the sensor data, and the state $x_t$ which includes the angle of roll, pitch, and yaw, respectively. The output of the code is the new set of particles $X_t$.  
  
According to the PF approach, the prediction considers only the motion information of the sensor. Basically, the pose of the moving node $x_t$ can be predicted using internal robot motion information such as motion command or odometer. Specifically, the internal motion information of a moving node, $u_{t}$, refers to the velocity command for surge, sway, heave, roll, pitch, and yaw motion respectively. Moreover, the motion model of the moving node formulates the
pose transition mathematically. The correction procedure calculates the degree of certainty signifying whether the predicted location indicates true location. Next, it resamples the probable location from the predicted location based on the assumption made. The certainty of each predicted location is calculated by comparing the real sensor data with the expected sensor data which will be resulted in if the moving node is located at
the predicted pose.
Resampling aims at  eliminating  the degeneracy which might result when significant weight is concentrated on only one particle after some time steps. It solves this issue by discarding particles with negligible weights and enhances ones with larger weights. 

\section{Simulation Results}

In this section, we apply  particle filter
method using the TDOA algorithm to enhance the localization accuracy. First, we set some basic parameters to the simulation scenario which are presented in Table 1. 

\begin{table}[h!] \label{simulation}
\caption{Simulation Parameter}
\normalsize
\begin{center}
\begin{tabular}{|c|c|c|}\hline
Symbol & Meaning & Value \\\hline
N & Number of Particles & 50  \\\hline
Q & Measurement Variance & 1 \\\hline
R & Process Variance & 3  \\\hline
$V_{max}$ & Maximum Velocity &5 $\frac{m}{s}$ \\\hline
$V_{min}$ & Minimum Velocity &1 $\frac{m}{s}$  \\\hline
$N_{x}$ & Length of x augmentation & 10  \\\hline
$N_{a}$ & Number of Anchors & 6  \\\hline
$N_{th}$ & Effective particle &10   \\\hline

\end{tabular}
\end{center}
\end{table}   

The localization performance is evaluated by the estimation
error in which the root mean square error (RMSE) is used to study the estimator's accuracy. The RMSE is defined as follows

\begin{equation}\label{sim}
RMSE = \sqrt{\frac{1}{T}\sum_{i}^{T}(x_{k}-\hat x_{k})^{2}}
\end{equation}

where $T$ is the number of measurements (time steps), $x_k$ is the real
position of a mobile node and $\hat x_k$ is the estimation of the
position.
The resampling scheme used is Residual resampling since it provides lower conditional variance for all configurations
of the weights \cite{douc2005comparison}. At one round, we use the RMSE to evaluate the
localization accuracy. We evaluate the localization accuracy
by the mean and variance of RMSE after many
rounds. As a comparison, we present the result of the other  filter algorithms,
including the Extended Kalman filter (EKF), the Unscented Kalman filter (UKF), and the particle filter (PF) based on TDOA measurement.
The values of the mean and variance of the different filtering algorithms can be found in Table 2.
\begin{table}[h!] \label{simulation}
\caption{Mean and Variance of Different Algorithms}
\normalsize
\begin{center}
\begin{tabular}{|c|c|c|}\hline
Filter & Mean & Variance \\\hline
EKF & 4.411 & 7.1947 \\\hline
UKF & 4.0583 & 1.6041\\\hline
PF &3.433  & 0.69623  \\\hline
PF-TDOA & 3.1624 &0.67719 \\\hline

\end{tabular}
\end{center}
\end{table} 
 
It can be noticed from Table 2  that the PF-TDOA approach has the least
mean, which indicates that the algorithm always has the best localization accuracy in comparison to the others.  

Figures 5 through 8 below demonstrate the accuracy of the proposed PF-TDOA approach in comparison to the other methods. In Figure 5, the capability of the PF approach to predict and correct the state of interest based on the observed sensor data can be inferred. Actually, each particle in the PF represents a state which translates into the moving node location in the case of TDOA localization. The PF-TDOA approach uses both the motion information as well as the distance estimates from the anchor nodes to be capable of providing a higher accuracy and enhanced robustness.  
\begin{figure}[h!]
\centering
\includegraphics[width=0.6\textwidth,clip]{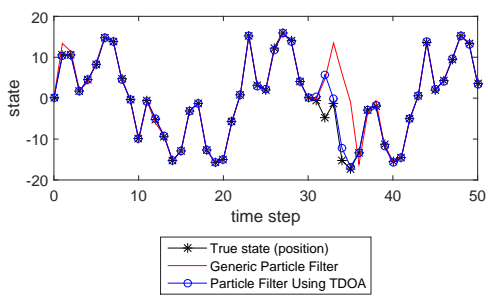}
\caption{Filter Estimates vs. True State}
\label{system}
\end{figure}

In Figure 6, the TDOA-PF approach is proven to have the lowest mean of estimation in comparison to the other

In Figure 7, the moving node follows a randomly generated path which is shown in blue. It can be easily noticed that the TDOA-PF approach closely depicts the actual trajectory of the moving node as it moves along the x and y coordinates in an open environment; hence, it is the most optimum. Actually, the fusion between the PF method and TDOA localization results in a reduced uncertainty region as illustrated in Figure 8.      
\begin{figure}[h]
\centering
\includegraphics[width=0.6\textwidth,clip]{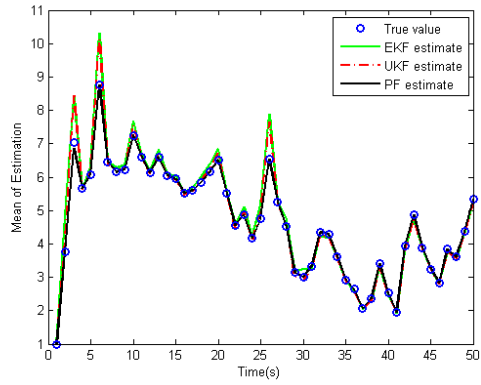}
\caption{Comparison between the performance of the PF, EKF and UKF filters}
\label{system}
\end{figure}

\begin{figure}[h!]
\centering
\includegraphics[width=0.6\textwidth,clip]{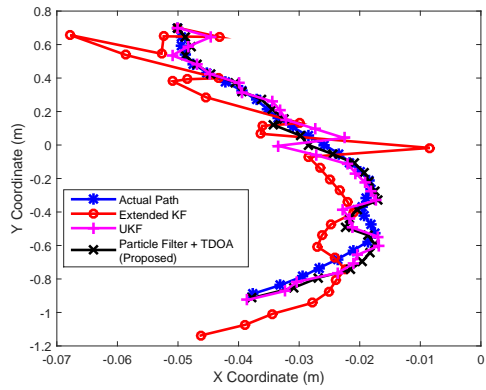}
\caption{Comparison between the performance of EKF, UKF and PF in TDOA localization.}
\label{system}
\end{figure}

\begin{figure}[h!]
\centering
\includegraphics[width=0.6\textwidth,clip]{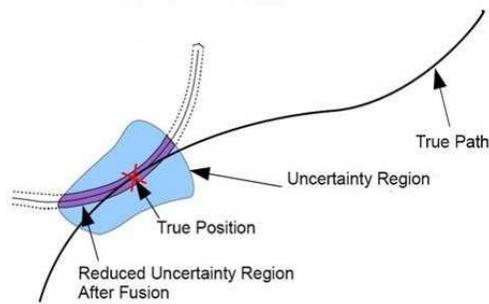}
\caption{Illustration of the TDOA-PF performance.}
\label{system}
\end{figure}

Next, we have evaluated the localization accuracy of both the general particle filter and that which utilizes the TDOA approach by assigning the simulation parameters with different values. Our goal is to find which parameters have
a more significant influence on the localization accuracy and robustness. 
\subsection{Number of Particles}
The
localization accuracy increases when the amount of particles
increase at most conditions. This result
actually conforms to the feature of the particle filter. 
\subsection{Number of Anchors}
The localization accuracy improves when increasing the number of anchors at
all algorithms. 
\subsection{Number of Iterations}

The particle filter uses the iterative process to
eliminate the noise in system and measurement. As a matter of fact, 
as the number of iterations increases, the higher the accuracy that can be achieved in the generic system state. However, in our system model, the
mobile node has to employ a random velocity when it moves forward in a time segment. Therefore, this might result in the  accumulation of error which leads to some localization errors.  
\section{Conclusion}
In this chapter, a technique for enhanced localization of moving nodes in Wireless Sensor Networks (WSNs) has been presented. The proposed technique is based on the use of Time Difference of Arrival (TDOA) along with Particle Filter (PF) method which is capable of accurate detection and enhanced localization since each particle in the PF represents a state which translates into the moving node location in the case of TDOA localization. The combined TDOA-PF technique utilizes  the internal motion information of the moving node as well as the distance estimates that result from the interaction between the anchor nodes. The performance of the TDOA-PF is compared to the other existing methods used for mobile node localization. Indeed, the simulation results proved that the proposed approach outperforms the other mentioned techniques in terms of the Root Mean Square Error (RMSE). \   
%+Bibliography
\footnotesize
\bibliographystyle{IEEEtran}
\bibliography{RefSetPIRMC20132}

% Generated by IEEEtran.bst, version: 1.14 (2015/08/26)
\begin{thebibliography}{10}
\providecommand{\url}[1]{#1}
\csname url@samestyle\endcsname
\providecommand{\newblock}{\relax}
\providecommand{\bibinfo}[2]{#2}
\providecommand{\BIBentrySTDinterwordspacing}{\spaceskip=0pt\relax}
\providecommand{\BIBentryALTinterwordstretchfactor}{4}
\providecommand{\BIBentryALTinterwordspacing}{\spaceskip=\fontdimen2\font plus
\BIBentryALTinterwordstretchfactor\fontdimen3\font minus
  \fontdimen4\font\relax}
\providecommand{\BIBforeignlanguage}[2]{{%
\expandafter\ifx\csname l@#1\endcsname\relax
\typeout{** WARNING: IEEEtran.bst: No hyphenation pattern has been}%
\typeout{** loaded for the language `#1'. Using the pattern for}%
\typeout{** the default language instead.}%
\else
\language=\csname l@#1\endcsname
\fi
#2}}
\providecommand{\BIBdecl}{\relax}
\BIBdecl

\bibitem{shubair2015vivo}
R.~M. Shubair and H.~Elayan, ``In vivo wireless body communications:
  State-of-the-art and future directions,'' in \emph{Antennas \& Propagation
  Conference (LAPC), 2015 Loughborough}.\hskip 1em plus 0.5em minus 0.4em\relax
  IEEE, 2015, pp. 1--5.

\bibitem{4407221}
A.~Boukerche, H.~Oliveira, E.~Nakamura, and A.~Loureiro, ``Localization systems
  for wireless sensor networks,'' \emph{Wireless Communications, IEEE},
  vol.~14, no.~6, pp. 6--12, December 2007.

\bibitem{Ilyas:2004:HSN:983182}
M.~Ilyas, I.~Mahgoub, and L.~Kelly, \emph{Handbook of Sensor Networks: Compact
  Wireless and Wired Sensing Systems}.\hskip 1em plus 0.5em minus 0.4em\relax
  Boca Raton, FL, USA: CRC Press, Inc., 2004.

\bibitem{mirebrahim2009monte}
H.~Mirebrahim and M.~Dehghan, ``Monte carlo localization of mobile sensor
  networks using the position information of neighbor nodes,'' in \emph{Ad-Hoc,
  Mobile and Wireless Networks}.\hskip 1em plus 0.5em minus 0.4em\relax
  Springer, 2009, pp. 270--283.

\bibitem{kulaib2011overview}
A.~Kulaib, R.~Shubair, M.~Al-Qutayri, and J.~W. Ng, ``An overview of
  localization techniques for wireless sensor networks,'' in \emph{Innovations
  in Information Technology (IIT), 2011 International Conference on}.\hskip 1em
  plus 0.5em minus 0.4em\relax IEEE, 2011, pp. 167--172.

\bibitem{5450293}
T.~Sathyan and M.~Hedley, ``Evaluation of algorithms for cooperative
  localization in wireless sensor networks,'' in \emph{Personal, Indoor and
  Mobile Radio Communications, 2009 IEEE 20th International Symposium on}, Sept
  2009, pp. 1898--1902.

\bibitem{hightower2001location}
J.~Hightower and G.~Borriello, ``Location systems for ubiquitous computing,''
  \emph{Computer}, no.~8, pp. 57--66, 2001.

\bibitem{wang2010survey}
J.~Wang, R.~K. Ghosh, and S.~K. Das, ``A survey on sensor localization,''
  \emph{Journal of Control Theory and Applications}, vol.~8, no.~1, pp. 2--11,
  2010.

\bibitem{najibi2013localization}
M.~Najibi, ``Localization algorithms in a wireless sensor network using
  distance and angular data,'' Ph.D. dissertation, Applied Sciences: School of
  Engineering Science, 2013.

\bibitem{shubair2007setup}
R.~M. Shubair, M.~Al-Qutayri, and J.~M. Samhan, ``A setup for the evaluation of
  music and lms algorithms for a smart antenna system.'' \emph{JCM}, vol.~2,
  no.~4, pp. 71--77, 2007.

\bibitem{al2004computationally}
E.~Al-Ardi, R.~Shubair, and M.~Al-Mualla, ``Computationally efficient
  high-resolution doa estimation in multipath environment,'' \emph{Electronics
  Letters}, vol.~40, no.~14, pp. 908--910, 2004.

\bibitem{al2005computationally}
E.~M. Al-Ardi, R.~M. Shubair, and M.~E. Al-Mualla, ``Computationally efficient
  doa estimation in a multipath environment using covariance differencing and
  iterative spatial smoothing,'' in \emph{Circuits and Systems, 2005. ISCAS
  2005. IEEE International Symposium on}.\hskip 1em plus 0.5em minus
  0.4em\relax IEEE, 2005, pp. 3805--3808.

\bibitem{al2003investigation}
E.~Al-Ardi, R.~Shubair, and M.~Al-Mualla, ``Investigation of high-resolution
  doa estimation algorithms for optimal performance of smart antenna systems,''
  2003.

\bibitem{samhan2006design}
J.~Samhan, R.~Shubair, and M.~Al-Qutayri, ``Design and implementation of an
  adaptive smart antenna system,'' in \emph{Innovations in Information
  Technology, 2006}.\hskip 1em plus 0.5em minus 0.4em\relax IEEE, 2006, pp.
  1--4.

\bibitem{al2003performance}
E.~Al-Ardi, R.~M. Shubair, and M.~E. Al-Mualla, ``Performance evaluation of
  direction finding algorithms for adapative antenna arrays,'' in
  \emph{Electronics, Circuits and Systems, 2003. ICECS 2003. Proceedings of the
  2003 10th IEEE International Conference on}, vol.~2.\hskip 1em plus 0.5em
  minus 0.4em\relax IEEE, 2003, pp. 735--738.

\bibitem{shubair2004robust}
R.~Shubair and A.~Al-Merri, ``Robust algorithms for direction finding and
  adaptive beamforming: performance and optimization,'' in \emph{Circuits and
  Systems, 2004. MWSCAS'04. The 2004 47th Midwest Symposium on}, vol.~2.\hskip
  1em plus 0.5em minus 0.4em\relax IEEE, 2004, pp. II--II.

\bibitem{nwalozie2013simple}
G.~Nwalozie, V.~Okorogu, S.~Maduadichie, and A.~Adenola, ``A simple comparative
  evaluation of adaptive beam forming algorithms,'' \emph{International Journal
  of Engineering and Innovative Technology (IJEIT)}, vol.~2, no.~7, 2013.

\bibitem{al2006direction}
E.~Al-Ardi, R.~Shubair, and M.~Al-Mualla, ``Direction of arrival estimation in
  a multipath environment: an overview and a new contribution,'' \emph{Applied
  Computational Electromagnetics Society Journal}, vol.~21, no.~3, 2006.

\bibitem{al2005direction}
M.~Al-Nuaimi, R.~Shubair, and K.~Al-Midfa, ``Direction of arrival estimation in
  wireless mobile communications using minimum variance distortionless
  response,'' in \emph{The Second International Conference on Innovations in
  Information Technology (IIT'05)}, 2005.

\bibitem{shubair2005robust}
R.~Shubair, ``Robust adaptive beamforming using lms algorithm with smi
  initialization,'' in \emph{Antennas and Propagation Society International
  Symposium, 2005 IEEE}, vol.~4.\hskip 1em plus 0.5em minus 0.4em\relax IEEE,
  2005, pp. 2--5.

\bibitem{bakhar2009eigen}
M.~Bakhar and D.~P. Hunagund, ``Eigen structure based direction of arrival
  estimation algorithms for smart antenna systems,'' \emph{IJCSNS International
  Journal of Computer Science and Network Security}, vol.~9, no.~11, pp.
  96--100, 2009.

\bibitem{shubair2005performance}
R.~Shubair and W.~Jessmi, ``Performance analysis of smi adaptive beamforming
  arrays for smart antenna systems,'' in \emph{Antennas and Propagation Society
  International Symposium, 2005 IEEE}, vol.~1.\hskip 1em plus 0.5em minus
  0.4em\relax IEEE, 2005, pp. 311--314.

\bibitem{alhajri2015hybrid}
M.~AlHajri, A.~Goian, M.~Darweesh, R.~AlMemari, R.~Shubair, L.~Weruaga, and
  A.~Kulaib, ``Hybrid rss-doa technique for enhanced wsn localization in a
  correlated environment,'' in \emph{Information and Communication Technology
  Research (ICTRC), 2015 International Conference on}.\hskip 1em plus 0.5em
  minus 0.4em\relax IEEE, 2015, pp. 238--241.

\bibitem{5432182}
S.~Zhang, J.~Cao, C.~Li-Jun, and D.~Chen, ``Accurate and energy-efficient
  range-free localization for mobile sensor networks,'' \emph{Mobile Computing,
  IEEE Transactions on}, vol.~9, no.~6, pp. 897--910, June 2010.

\bibitem{dellaert1999monte}
F.~Dellaert, D.~Fox, W.~Burgard, and S.~Thrun, ``Monte carlo localization for
  mobile robots,'' in \emph{Robotics and Automation, 1999. Proceedings. 1999
  IEEE International Conference on}, vol.~2.\hskip 1em plus 0.5em minus
  0.4em\relax IEEE, 1999, pp. 1322--1328.

\bibitem{handschin1970monte}
J.~Handschin, ``Monte carlo techniques for prediction and filtering of
  non-linear stochastic processes,'' \emph{Automatica}, vol.~6, no.~4, pp.
  555--563, 1970.

\bibitem{rui2001better}
Y.~Rui and Y.~Chen, ``Better proposal distributions: Object tracking using
  unscented particle filter,'' in \emph{Computer Vision and Pattern
  Recognition, 2001. CVPR 2001. Proceedings of the 2001 IEEE Computer Society
  Conference on}, vol.~2.\hskip 1em plus 0.5em minus 0.4em\relax IEEE, 2001,
  pp. II--786.

\bibitem{so2011source}
H.~C. So, ``Source localization: algorithms and analysis,'' \emph{Handbook of
  Position Location: Theory, Practice, and Advances}, pp. 25--66, 2011.

\bibitem{laaraiedh2010contributions}
M.~Laaraiedh, ``Contributions on hybrid localization techniques for
  heterogeneous wireless networks,'' Ph.D. dissertation, Universit{\'e} Rennes
  1, 2010.

\bibitem{4524057}
H.~C. So, Y.~T. Chan, and F.~K.~W. Chan, ``Closed-form formulae for
  time-difference-of-arrival estimation,'' \emph{IEEE Transactions on Signal
  Processing}, vol.~56, no.~6, pp. 2614--2620, June 2008.

\bibitem{boudhir2010new}
A.~A. Boudhir, B.~Mohamed, and B.~A. Mohamed, ``New technique of wireless
  sensor networks localization based on energy consumption,'' \emph{Int. J.
  Comput. Appl}, vol.~9, no.~12, pp. 25--28, 2010.

\bibitem{cheung2006constrained}
K.~W. Cheung, H.-C. So, W.-K. Ma, and Y.-T. Chan, ``A constrained least squares
  approach to mobile positioning: algorithms and optimality,'' \emph{EURASIP
  journal on applied signal processing}, vol. 2006, pp. 150--150, 2006.

\bibitem{zhou2011tracking}
F.~Zhou, Z.~Qin, C.~Xiao, S.~Li, W.~Jiang, and Y.~Wu, ``Tracking moving object
  via unscented particle filter in sensor network.'' \emph{International
  Journal of Digital Content Technology \& its Applications}, vol.~5, no.~12,
  2011.

\bibitem{1098986}
M.~Athans, R.~Wishner, and A.~Bertolini, ``Suboptimal state estimation for
  continuous-time nonlinear systems from discrete noisy measurements,''
  \emph{Automatic Control, IEEE Transactions on}, vol.~13, no.~5, pp. 504--514,
  Oct 1968.

\bibitem{882463}
E.~Wan and R.~Van Der~Merwe, ``The unscented kalman filter for nonlinear
  estimation,'' in \emph{Adaptive Systems for Signal Processing,
  Communications, and Control Symposium 2000. AS-SPCC. The IEEE 2000}, 2000,
  pp. 153--158.

\bibitem{978374}
M.~Arulampalam, S.~Maskell, N.~Gordon, and T.~Clapp, ``A tutorial on particle
  filters for online nonlinear/non-gaussian bayesian tracking,'' \emph{Signal
  Processing, IEEE Transactions on}, vol.~50, no.~2, pp. 174--188, Feb 2002.

\bibitem{6263573}
N.~Y. Ko, T.~G. Kim, and Y.~S. Moon, ``Particle filter approach for
  localization of an underwater robot using time difference of arrival,'' in
  \emph{OCEANS, 2012 - Yeosu}, May 2012, pp. 1--7.

\bibitem{6817863}
J.~Bordoy, P.~Hornecker, F.~Hoflinger, J.~Wendeberg, R.~Zhang,
  C.~Schindelhauer, and L.~Reindl, ``Robust tracking of a mobile receiver using
  unsynchronized time differences of arrival,'' in \emph{Indoor Positioning and
  Indoor Navigation (IPIN), 2013 International Conference on}, Oct 2013, pp.
  1--10.

\bibitem{1236770}
P.~Djuric�, J.~H. Kotecha, J.~Zhang, Y.~Huang, T.~Ghirmai, M.~Bugallo, and
  J.~Miguez, ``Particle filtering,'' \emph{Signal Processing Magazine, IEEE},
  vol.~20, no.~5, pp. 19--38, Sep 2003.

\bibitem{douc2005comparison}
R.~Douc and O.~Capp{\'e}, ``Comparison of resampling schemes for particle
  filtering,'' in \emph{Image and Signal Processing and Analysis, 2005. ISPA
  2005. Proceedings of the 4th International Symposium on}.\hskip 1em plus
  0.5em minus 0.4em\relax IEEE, 2005, pp. 64--69.

\end{thebibliography}

%-Bibliography

\end{document}